\documentclass[aps,twocolumn,preprintnumbers,groupedaddress,nofootinbib]{revtex4-1}
\pdfoutput=1

\usepackage{tabu}
\usepackage[usenames,dvipsnames,table]{xcolor}
\usepackage{graphicx,amsmath,amssymb,amsthm,multirow,array,bm,bbm,esint}
\usepackage[mathscr]{eucal}
\usepackage[bbgreekl]{mathbbol}
\usepackage{epsf,amsfonts}
\usepackage{physics}
\usepackage{dsfont, mathtools}
\usepackage{slashed}
\usepackage{hyperref}
\usepackage{mathdots}

\usepackage[normalem]{ulem}

\hypersetup{
    pdfstartview={FitH},    
    pdftitle={Discovering the wormhole from quantum chaos and number theory in two-dimensional CFTs},    
    pdfauthor={Felix Haehl},     
    colorlinks=true,       
    linkcolor=blue,          
    citecolor=red,        
    filecolor=magenta,      
    urlcolor=blue           
}

\usepackage{tikz}

\begin{document}

\title{Euclidean wormholes in two-dimensional CFTs\\ from quantum chaos and number theory}

\author{Felix M.\ Haehl}
\affiliation{School of Mathematical Sciences, University of Southampton, SO17 1BJ, U.K.}
\author{Wyatt Reeves}
\author{Moshe Rozali}
\affiliation{Department of Physics and Astronomy, University of British Columbia, Vancouver, V6T 1Z1, Canada}

\begin{abstract}
We consider two-dimensional conformal field theories (CFTs), which exhibit a hallmark feature of quantum chaos: universal repulsion of energy levels as described by a regime of linear growth of the spectral form factor. This physical input together with modular invariance strongly constrains the spectral correlations and the subleading corrections to the linear growth. We show that these are determined by the Kuznetsov trace formula, which highlights an intricate interplay of universal physical properties of chaotic CFTs and analytic number theory. The trace formula manifests the fact that the simplest possible CFT correlations consistent with quantum chaos are precisely those described by a Euclidean wormhole in AdS${}_3$ gravity with [torus]$\times$[interval] topology.  For contrast, we also discuss examples of non-chaotic CFTs in this language.
\end{abstract}

\pacs{}

\maketitle

\section{Introduction}
\label{sec:intro}

Given the abundance of systems exhibiting statistical universalities referred to as quantum chaos, including black holes \cite{Cotler:2016fpe}, it is of obvious interest to study these universalities in some of the most important models for holography: two-dimensional conformal field theories (CFTs).

A prototypical example of a gravitational off-shell solution with two boundaries is the asymptotically AdS wormhole with  [torus]$\times$[interval] ($\mathbb{T}^2 \times I$) topology (c.f., \cite{Maldacena:2004rf}). The two-boundary CFT partition function associated with such a geometry was studied by Cotler-Jensen in \cite{Cotler:2020ugk}. In particular it was found that, interpreted as a spectral form factor (SFF), this wormhole amplitude exhibits random matrix universality in the near-extremal long-time limit and hence describes correlations in what can be called a chaotic CFT. The same authors also initiated a bootstrap approach, which aims to derive the wormhole amplitude from minimal assumptions about CFTs \cite{Cotler:2020hgz}. This idea has benefited enormously from expressing CFT quantities in a manifestly modular invariant basis \cite{Benjamin:2021ygh}. In particular, Di Ubaldo-Perlmutter derived the wormhole amplitude using spectral theory and making only a few natural assumptions about the pairing of eigenvalues and eigenfunctions of the Laplacian on the fundamental domain \cite{DiUbaldo:2023qli}. Similarly, in \cite{Haehl:2023wmr} we show how the wormhole amplitude can be discovered by demanding random matrix universality (for a single CFT) in the appropriate limit, independently in each spin sector.

The study of quantum chaos in 2d CFTs is complicated by the fact that these systems enjoy an enormous amount of symmetry, which rigidly dictates much of the spectrum. To discuss chaos,\footnote{This is in the spirit first introduced in \cite{Pollack:2020gfa}, where we envision an effective coarse-graining of states in a single CFT, similar to the coarse-graining underlying quantum statistical mechanics of closed unitary systems.} one needs to focus on a superselection sector by: $(i)$ removing Virasoro descendants, $(ii)$ discarding states that are images under modular transformations of the non-chaotic `censored' states, i.e., those states with conformal weights satisfying $\text{min}(h,\bar{h}) \leq \frac{c-1}{24}$ \cite{Keller:2014xba,Schlenker:2022dyo}. This can always be achieved in a way that preserves modular invariance \cite{Benjamin:2021ygh}. The result of this procedure is the `fluctuating' part of the spectrum of primary states, describing oscillations around the average physical density of states, i.e., statistical fluctuations in the dual BTZ black hole microstate spectrum. We refer to its partition function as $\widetilde{Z}_\text{P}$. Finally, $(iii)$ we focus on the partition function $\widetilde{Z}_\text{P}^m$ of states with definite spin $m=h-\bar{h}$.

Quantum chaos refers to the fact that states with nearby energies are correlated in a universal way that encodes the repulsion of energy levels. E.g., if $\widetilde{\rho}^{\,m}_\text{P}(E)$ denotes the density of spin $m$ operators counted by $\widetilde{Z}^m_\text{P}$, random matrix universality is the statement that
\begin{equation}
    \big\langle \widetilde{\rho}^{\,m_1}_\text{P}(E_1) \widetilde{\rho}^{\,m_2}_\text{P}(E_2) \big\rangle \sim - \frac{\delta_{m_1m_2}}{\pi^2 |\omega|^2}   \quad ( \omega\ll E_k - E_{m_k} \ll 1) \,,
\end{equation}
where $\omega \equiv E_1 - E_2$ and $E_{m} \equiv 2\pi ( m - \frac{1}{12} )$ is the lowest energy in the spectrum of $\widetilde{Z}^m_\text{P}$. In the time domain, the effect of this term is a linear growth of the SFF, often referred to as a {\it ramp}: we place two copies of the CFT on tori with modular parameters $\tau_k = x_k + i y_k$ and analytically continue $y_{1,2} \rightarrow \beta \pm iT$. In the near-extremal $\beta \gg 1$ and late-time $T\gg \beta$ limit the linear (in $T$) ramp follows from the following (Euclidean) behavior:\footnote{Our normalization of the wormhole amplitude differs by a factor of 2 from \cite{Cotler:2020ugk}. This is in order to describe the GOE universality class, c.f., \cite{Yan:2023rjh} and comments in \cite{DiUbaldo:2023qli}.}
\begin{equation}
\label{eq:ramp}
 \big{\langle} \widetilde{Z}_\text{P}^{m_1}(y_1)  \widetilde{Z}_\text{P}^{m_2}(y_2) \big{\rangle}= \frac{\delta_{m_1m_2}}{\pi} \frac{y_1 y_2}{y_1+y_2}\, e^{-2\pi |m_1|(y_1+y_2)} + \ldots 
\end{equation}
up to subleading terms in the limit $y_k \gg 1$ with $\frac{y_1}{y_2}$ held fixed. We refer to these asymptotics as a {\it `bare ramp'}, as it encodes nothing further than the minimum amount of information that follows universally from quantum chaos.\footnote{On first inspection, our `linear' ramp grows as $T^2$. This is due to the modular invariant construction of $\widetilde{Z}_\text{P}$, which introduces a spurious factor of $\sqrt{y_1y_2} \sim T$. The physical SFF follows after removing this factor. See, e.g., \cite{Benjamin:2021ygh,Haehl:2023tkr}.} The presence of this ramp is the defining feature of quantum chaos and it is the main assumption we make about the spectrum of the CFT. We revisit the following question: assuming only a linear ramp \eqref{eq:ramp} for every spin sector, how can we make it consistent with modular invariance and what do we learn about the SFF of the full theory? The elegant answer is that certain subleading terms need to be added to the `bare ramp' in order to restore modular invariance  \cite{DiUbaldo:2023qli}. These terms are dictated by general symmetry considerations, which we phrase as stringent requirements rooted in analytic number theory. We quantify a minimality assumption about subleading corrections to the `bare ramp', which leads to the $\mathbb{T}^2\times I$ gravitational constrained instanton of \cite{Cotler:2020ugk}.

\section{Spectral theory and Kuznetsov trace formula}
\label{sec:kuznetsov}

The central result, which we use to find the subleading terms minimally completing the `bare' ramp into a modular invariant amplitude, is the Kuznetsov trace formula. The formula connects the {\it spectral} theory of the Laplacian on the fundamental domain ${\cal F} = \mathbb{H}/SL(2,\mathbb{Z})$ to {\it geometric} Poincar\'{e} series. To set up notation, we briefly review the spectral part of this formalism. The spectrum of the Laplacian on ${\cal F}$ has a continuous and a discrete part. A basis of modular invariant parity-invariant functions are the following Eisenstein series (labelled by $\alpha \in \mathbb{R}$) and Maass cusp forms (labelled by $n\in \mathbb{Z}_+$):\footnote{We focus on CFTs with parity symmetry, where we can project onto the parity-even superselection sector that only requires parity-even cusp forms. See SM \ref{app:odd} for a discussion of odd forms.}
\begin{align}
\label{eq:basis}
        E_{\frac{1}{2} + i \alpha} &=  \sum_{m\geq 0} \cos(2\pi m x)\, \frac{(2-\delta_{m,0})}{\Lambda(-i\alpha)} \, a_m^{(\alpha)}  \sqrt{y} K_{i\alpha}(2\pi m y)  \,, \nonumber\\ \nu_{n} &= \sum_{m\geq 1} \cos(2\pi m x)\, a_m^{(n)} \, \sqrt{y} K_{iR_n} (2\pi m y) \,,
\end{align}
where $\Lambda(\frac{s}{2}) \equiv \Lambda(\frac{1-s}{2}) \equiv  \pi^{-s/2} \Gamma(s/2) \zeta(s)$ is the completed Riemann $\zeta$-function.
The eigenvalues of the Laplacian are $\frac{1}{4} + \alpha^2$ and $\frac{1}{4} + R_n^2$, respectively, where the $R_n>0$ are sporadic, Poisson distributed numbers whose density grows linearly with the value of $R_n$. 
The Fourier coefficients of Eisenstein series are $a_m^{(\alpha)} \equiv  2|m|^{-i\alpha}\sigma_{2i\alpha}(|m|)$.\footnote{The divisor sum is $\sigma_z(m) \equiv \sum_{d|m} d^z$, implying $a_m^{(\alpha)} = a_m^{(-\alpha)}$.}
The Fourier coefficients of the cusp forms, $a_m^{(n)} \equiv a_{-m}^{(n)}$, are again erratic discrete numbers, which are statistically distributed according to known distributions (we normalize $a_1^{(n)}=1$). For instance, $a_p^{(n)}\in (-2,2)$ are independently distributed for different primes $p$ with a distribution that approaches a Wigner semi-circle centered at 0 for large $p$; the coefficients for non-prime spins can be constructed from those with prime spin via the fact that $\nu_n(\tau)$ are eigenfunctions of Hecke operators. This random but highly constrained structure and its statistical properties is called {\it arithmetic chaos} \cite{bolteArithmeticalChaosViolation1992,sarnakthesis}.

Modular invariant quantities of sufficiently fast decay at the cusp $y\rightarrow \infty$ can be expanded in the  basis of Eisenstein series and cusp forms \cite{Terras2013HarmonicAO}. In particular: 
\begin{gather}
 \widetilde{Z}_\text{P}(\tau) = \langle \widetilde{Z}_\text{P} \rangle + \int_{\mathbb{R}} \frac{d\alpha}{4\pi} \, z_{\frac{1}{2}+i\alpha} \, E_{\frac{1}{2}+i\alpha}(\tau) + \sum_{n\geq 1} z_n \, \nu_{n}(\tau) \nonumber \\
z_{\frac{1}{2}+i\alpha} =  \big( \widetilde{Z}_\text{P},\, E_{\frac{1}{2}+i\alpha} \big) \,,\quad
z_n = \frac{(\widetilde{Z}_\text{P},\,\nu_{n})}{|\!|\nu_{n}|\!|^2} \,,
\label{eq:specDecomp}
\end{gather}
where $(\,\cdot \, , \, \cdot \,)$ denotes the Petersson ($L^2$-)inner product on ${\cal F}$. The first term in \eqref{eq:specDecomp} is the spectral average of $\widetilde{Z}_\text{P}$, i.e., its overlap with $\nu_0 \equiv 1$; it vanishes by construction.
Since the assumption of quantum chaos should be unconstrained by symmetries, it is natural to expand the SFF $\langle \widetilde{Z}_\text{P}(\tau_1) \widetilde{Z}_\text{P}(\tau_2 )\rangle$ in (two copies of) the manifestly modular invariant basis \eqref{eq:basis}. The coefficients in such an expansion will be {\it correlators} of the overlap coefficients $z_{\frac{1}{2}+i\alpha}$ and $z_n$. Such correlators should be understood in the sense of coarse-graining: they quantify the correlations in the spectrum as a function of the spectral parameters $\alpha$ and $n$ (which is related to the more common correlations in nearby energy windows by an integral transform).

The central result we use from analytic number theory is the following (see \cite{bruggeman,kuznetsov1980petersson}, also \cite{motohashi1997}):

\begin{widetext}
\noindent{\it {\bf Theorem (Kuznetsov):} Let $h(\alpha)$ be an even function, which is holomorphic and sufficiently fast decaying in an appropriate region of the complex plane.\footnote{ More precisely, it is required that $h(\alpha)$ is regular for $|\text{Im}(\alpha)|\leq \frac{1}{2}+\delta$ and in that region $|h(\alpha)| \ll (1+|\alpha|)^{-2-\delta}$ for some $\delta > 0$.} Then, for $|m_k|\geq 1$:}
\begin{equation}
\label{eq:kuznetsov}
    \begin{split}
        &\int_{\mathbb{R}} \frac{d\alpha}{4\pi} \, \frac{a_{m_1}^{(\alpha)} a_{m_2}^{{(\alpha)}}}{2L^{(2\alpha)}_E(1)} \, h(\alpha) +  \sum_{n\geq 1} \frac{a_{m_1}^{(n)}a_{m_2}^{(n)}}{L_{\nu\times\nu}^{(n)}(1)} \, h(R_n)  = \frac{\delta_{m_1m_2}}{\pi} \int_{\mathbb{R}}\frac{d\alpha}{4\pi}  \,\alpha  \tanh(\pi \alpha) h(\alpha) + {\cal G}^+_{m_1m_2}\!\!\!\!
    \end{split}
\end{equation}
{\it where the final term mixes spin sectors and is given in terms of Kloosterman sums:}\,\footnote{Recall $S(a,b;c) \equiv \sum_\ell e^{2\pi i(a \ell + b \bar\ell)/c}$ where the sum is over $1\leq \ell \leq c$ with gcd$(\ell,c) = 1$. Here, $\bar\ell$ is such that $\ell \bar\ell = 1 (\text{mod } c)$.}
\begin{equation}
\label{eq:Kdef}
\begin{split}
    {\cal G}^+_{m_1m_2} &=  2i\, \sum_{c\geq 1} \frac{S(|m_1|,|m_2|;c)}{c} \int_{\mathbb{R}} \frac{d\alpha}{4\pi} \, \frac{\alpha\,h(\alpha)}{\cosh(\pi \alpha)} \, J_{2i\alpha}\left( \frac{4\pi \sqrt{|m_1m_2|}}{c} \right)  \\
    &\quad + \frac{4}{\pi} \,\sum_{c\geq 1} \frac{S(|m_1|,-|m_2|;c)}{c} \int_{\mathbb{R}}\frac{d\alpha}{4\pi} \, \alpha \, h(\alpha)\sinh(\pi \alpha) \, K_{2i\alpha}\left( \frac{4\pi \sqrt{|m_1m_2|}}{c} \right)
\end{split}
\end{equation}
\end{widetext}

The l.h.s.\ of the theorem involves Rankin-Selberg $L$-functions for Eisenstein series and cusp forms. These are generalizations of the Riemann $\zeta$-function, which can be defined as series over $m$ involving the Fourier coefficients:
\begin{equation}
L_E^{(\alpha)}(s) = \frac{1}{2}\sum_{m\geq 1} \frac{a_m^{(\alpha)}}{m^s} \,,\quad 
L_{\nu\times\nu}^{(n)}(s) = \frac{\zeta(2s)}{\zeta(s)} \sum_{m\geq 1} \frac{\big(a_{m}^{(n)}\big)^2}{m^s} 
\end{equation}
for Re$(s)>1$. Similar to the Riemann $\zeta$-function the $L$-functions admit a meromorphic continuation to the entire complex plane. 
Evaluated at $s=1$ these simplify as follows \cite{blomer2019symplectic}:
\begin{equation}
\label{eq:Leval1}
\begin{split}
 L_E^{(2\alpha)}(1) &= |\zeta(1+2i\alpha) |^2 \equiv \cosh(\pi \alpha) |\Lambda(i\alpha)|^2\,,\\
 L_{\nu\times\nu}^{(n)}(1) &= 8 \cosh(\pi R_n) |\!|\nu_n|\!|^2 \,.
\end{split}
\end{equation}

The result \eqref{eq:kuznetsov} is a trace formula as it computes the trace of certain Hecke operators. Their eigenvalues are proportional to the Fourier coefficients and the trace formula contains these in a pattern with correlated (`diagonal') eigenvalues $\alpha$ and $n$. The $L$-functions provide the number theoretic kernels for the traces of Hecke operators. The r.h.s.\ of the trace formula should be understood in a geometric sense: it originates from computing Fourier coefficients of a Poincar\'{e} series. The first term, which is diagonal in spin, arises from translations. The spin-mixing second term arises from all other $SL(2,\mathbb{Z})$ transformations. Clearly quantum chaos, being an independent feature of fixed spin sectors, should be encoded in the first term.

\section{Minimal modular completion and spectral decomposition of the ramp}
\label{sec:specdecomp}

Let us now apply the Kuznetsov trace formula to the universal ramp in chaotic CFTs. First note that, if 
\begin{equation}
\label{eq:hdef}
    h(\alpha) = 4 \sqrt{y_1} K_{i\alpha}(2\pi |m_1| y_1) \sqrt{y_2} K_{i\alpha}(2\pi |m_2| y_2) \, g(\alpha)
\end{equation} 
with $g$ independent of $y_k$ and spins, then the l.h.s.\ of \eqref{eq:kuznetsov} involves fixed spin components of Eisenstein series and cusp forms, thus providing an $SL(2,\mathbb{Z})$ spectral decomposition for the trace of a product of modular invariant functions.\footnote{ This is sometimes called `pre-Kuznetsov formula', see \eqref{eq:prekuz}.} 

We first consider the {\it simplest possible function} that furnishes  such a spectral decomposition: a constant,
\begin{equation}
\label{eq:hwh}
 g^\text{(wh)}(\alpha) = 1 \,,
\end{equation}
and denote by $h^\text{(wh)}(\alpha)$ the corresponding function \eqref{eq:hdef}.
Using a standard Bessel function integral, the spin-diagonal first term on the r.h.s.\ of \eqref{eq:kuznetsov} is the `bare' ramp:
\begin{equation}
\label{eq:WHcorrelations}
\begin{split}
&\frac{\delta_{m_1m_2}}{\pi} \int_{\mathbb{R}} \frac{d\alpha}{4\pi} \, \alpha  \tanh(\pi \alpha) \, h^\text{(wh)}(\alpha)
\\&\qquad\qquad  = \frac{\delta_{m_1m_2}}{\pi} \frac{y_1 y_2}{y_1+y_2}\, e^{-2\pi |m_1|(y_1+y_2)} 
\end{split}
\end{equation}
Note that only for the choice \eqref{eq:hwh} will \eqref{eq:WHcorrelations} be the `bare' ramp with all corrections subsumed in ${\cal G}_{m_1m_2}^\text{+(wh)}$, thus realizing the quantum chaos assumption in a {\it minimal} way. This has several immediate consequences. First, the l.h.s.\ of the trace formula must provide a spectral decomposition of the ramp. We can simply read off the coefficients of this decomposition from \eqref{eq:kuznetsov} and \eqref{eq:Leval1}:
\begin{equation}
\label{eq:summaryRamp}
    \begin{split}
    \big\langle  z_{\frac{1}{2}+i\alpha_1} \, z_{\frac{1}{2}+i\alpha_2}  \big\rangle_{\text{(wh)}} 
    &= \frac{1}{2\cosh(\pi \alpha_1)} \times 4\pi \delta(\alpha_1-\alpha_2) \,,\\
    \big\langle z_{n_1} \, z_{n_2} \big\rangle_{\text{(wh)}} &=  \frac{1}{2\cosh(\pi R_{n_1})} \, \frac{1}{|\!|\nu_{n_1}|\!|^2}\times\delta_{n_1n_2} \,.
 \end{split}
\end{equation}
These are indeed known expressions: \eqref{eq:summaryRamp} has recently been identified as the spectral decomposition of the $\mathbb{T}^2\times I$ wormhole in AdS$_3$ gravity \cite{DiUbaldo:2023qli}.\footnote{We also thank S.\ Collier for private communication on this result.}
Further, in \cite{Haehl:2023wmr} we derive the same result from statistical considerations by demanding consistency across spin sectors of the quantum chaos assumption in a minimal way. Here, we got this result as an immediate consequence of the Kuznetsov trace formula applied to the ramp. To summarize, we write this minimal application of the trace formula, which describes the wormhole amplitude as follows:
\begin{widetext}
\begin{equation}
\label{eq:fnormRel}
\begin{split}
   &\int_\mathbb{R}\frac{d\alpha}{4\pi} \frac{1}{2\cosh(\pi\alpha)} \, E^{m_1}_{\frac{1}{2}+i\alpha}(y_1) \, E^{m_2}_{\frac{1}{2}+i\alpha}(y_2)
   +  \sum_{n \geq 1} \frac{1}{2\cosh(\pi R_n) }  \frac{\nu_{n}^{m_1}(y_1)}{|\!|\nu_{n}|\!|} \frac{\nu_{n}^{m_2}(y_2)}{|\!|\nu_{n}|\!|} = \frac{\delta_{m_1m_2}}{\pi} \frac{y_1 y_2}{y_1+y_2}\, e^{-2\pi |m_1|(y_1+y_2)}  + {\cal G}_{m_1m_2}^{+\text{(wh)}}
\end{split}
\end{equation}
\end{widetext}
where $E^{m}_{\frac{1}{2}+i\alpha}$ and $\nu_n^m$ are the spin $m$ components of the basis functions \eqref{eq:basis}. The cusp form norms appearing in denominators ensure that the expansion is w.r.t.\ an orthonormal basis. 

The first term on the l.h.s.\ of \eqref{eq:fnormRel} can be shown to generate the linear ramp for spin 0; for $|m_k|\geq 1$ it contributes a subleading term of the form $\sim \delta_{m_1m_2} \sqrt{y_1y_2/(y_1+y_2)}\, e^{-2\pi |m_1|(y_1+y_2)}$, c.f. \cite{Haehl:2023tkr}. The second term on the l.h.s.\ generates the linear ramp for spins $|m_k|>0$ (plus further subleading terms), c.f.\ \cite{DiUbaldo:2023qli,Haehl:2023wmr}. 

Consider now the r.h.s.\ of \eqref{eq:fnormRel}: it tells us precisely which terms are required in order for the `bare' ramp (first term) to be made consistent with modular invariance and the trace formula, while organizing them in a useful fashion.
From the comments above, we must expect that ${\cal G}_{m_1m_2}^{+\text{(wh)}}$ matches the subleading corrections found in the gravity calculation of \cite{Cotler:2020ugk}. To see that this is the case, we begin with the second line of \eqref{eq:Kdef}, which is elementary and yields, for $h^\text{(wh)}(\alpha)$:
\begin{equation}
\label{eq:GwhOpposite}
\begin{split}
{\cal G}_{m_1m_2}^{+\text{(wh)}} &\supset  
\sum_{c\geq 1} \frac{S(|m_1|,-|m_2|;c)}{c^2B_c} 
\, e^{-2\pi(|m_1|y_1+|m_2|y_2) B_c} 
\end{split}
\end{equation}
where $B_c \equiv \big(1+\frac{1}{c^2y_1y_2}\big)^{1/2}$. It is immediately clear that this is subleading compared to the ramp \eqref{eq:WHcorrelations} in the late time near-extremal limit. Further, \eqref{eq:GwhOpposite} indeed matches the subleading terms found in the gravity analysis of the wormhole \cite{Cotler:2020ugk} in the case where $\text{sgn}(m_1m_2)= -1$. The first line of \eqref{eq:Kdef} is more complicated, but can be shown to match the gravity result when $\text{sgn}(m_1m_2)= 1$ (see SM \ref{app:subleading})\footnote{Our analysis of only the parity even spectrum corresponds to adding up the result of \cite{Cotler:2020ugk} for same and for opposite sign spins. Variations of this analysis are described in the SM.} -- ultimately because the latter implements a Poincar\'{e} sum over certain modular invariant seed functions, which is precisely what the r.h.s.\ of the trace formula captures.

On the one hand, these subleading terms are rather subtle: they contain all the erratic information about `arithmetic chaos' exhibited by the infinite set of cusp forms in just the right way (c.f., \cite{Haehl:2023wmr}), reorganizing it cleanly into Kloosterman sums. On the other hand, the subleading terms are very simple -- in fact as simple as they can possibly be: they complete the `bare' ramp in all spin sectors, i.e., the fundamental input required by the assumption of quantum chaos, into a quantity that is consistent with conformal symmetry in the minimal way. Indeed, the choice $g^\text{(wh)}(\alpha)$ leading to the wormhole amplitude was manifestly the minimal option that would yield any modular invariant spectral decomposition at all. This simplicity of the gravity amplitude was first emphasized in \cite{DiUbaldo:2023qli} and was dubbed as MaxRMT (`maximal random matrix theory') principle.\footnote{Their discussion rests on similar assumptions and minimality requirements realized by the wormhole amplitude (in particular the diagonal pairing of eigenvalues and eigenfunctions of \eqref{eq:fnormRel}), but formalized in the context of the Gutzwiller trace formula.}

Note that our application of the trace formula did not use any input other than the assumption of universal level repulsion (linear ramp).\footnote{ The trace formula applies to objects with an underlying structure of Poincar\'{e} sums, see SM \ref{app:subleading}. This explains the consistency with the results of \cite{Cotler:2020ugk} and \cite{DiUbaldo:2023qli}.} In particular, the following features were already built into the mechanism of the trace formula and are hence identified as a natural and consistent starting point: $(i)$ the diagonal pairing of $SL(2,\mathbb{Z})$ eigenvalues; $(ii)$ the diagonality in spin at leading order for large $y_k$; $(iii)$ the fact that the correlations of overlap coefficients \eqref{eq:summaryRamp} had the same functional form in the continuous and discrete sectors; $(iv)$ the determination of subleading terms in the large $y_k$ limit.

\section{Examples without chaos}

\subsection{Narain CFTs}
\label{sec:plateau}

For contrast, and to usefully extend the applicability of the trace formula, we will now discuss a different application, which describes the SFF of an {\it integrable} ensemble of 2d CFTs; namely, we consider the Narain theories of $D$ free lattice bosons, which enjoy a $U(1)^{D}\times U(1)^{D}$ global symmetry \cite{Maloney:2020nni,Afkhami-Jeddi:2020ezh}. A dual description in terms of Chern-Simons theory has been further explored in \cite{Cotler:2020hgz} (see also \cite{Benjamin:2021wzr}). We will momentarily reproduce the associated $\mathbb{T}^2\times I$ amplitude from the trace formula, using a similar approach as for the wormhole in pure gravity.

The primary state counting partition function is 
\begin{equation}
    Z_{\text{P}(D)} =  y^{D/2} |\eta(x+iy)|^{2D} Z_\text{Narain} \,,
\end{equation}
where the prefactor removes Virasoro descendants of $D$ bosons in a modular invariant way. $Z_{\text{P}(D)}$ is amenable to spectral analysis without further modification \cite{Benjamin:2021ygh}. The $\mathbb{T}^2 \times I$ wormhole contribution to the SFF is \cite{Cotler:2020hgz,Collier:2021rsn}:
\begin{equation}
\begin{split}
    &\big\langle {Z}_{\text{P}(D)}^{m_1} {Z}_{\text{P}(D)}^{m_2} \big\rangle_\text{(wh)}  =  \delta_{m_1m_2} \,\frac{2\pi^{\frac{D}{2}}}{\Gamma\left(\frac{D}{2} \right)} \left( \frac{|m_1|}{y_1+y_2} \right)^{\frac{D-1}{2}}\\ &\qquad\qquad\qquad\qquad \times (y_1y_2)^{\frac{D}{2}} K_{\frac{D-1}{2}}\big(2\pi|m_1|(y_1+y_2)\big)\,.
\end{split}
\label{eq:CJplateau}
\end{equation}
The interpretation of \eqref{eq:CJplateau} is in terms of a universal {\it plateau}, which reflects the discreteness of the spectrum of this non-chaotic theory. The late-time limit of the SFF is a temperature-dependent constant:
\begin{equation}
\label{eq:SFFnarain}
    \begin{split}
    &(y_1y_2)^{-\frac{D}{2}} \big\langle {Z}_{\text{P}(D)}^{m_1}(y_1) \,{Z}_{\text{P}(D)}^{m_2} ( y_2) \big\rangle \big{|}_{y_{1,2} \rightarrow \beta \pm iT} \\
    &\quad \sim \text{const.} \times  \beta^{-\frac{D}{2}} \, e^{-4\pi|m_1| \beta} \,\delta_{m_1m_2}\qquad (T \gg \beta) \,.
    \end{split}
\end{equation}
We can obtain this expression from the trace formula by choosing a function $g(\alpha)$ that generalizes the simplest option \eqref{eq:hwh}:
\begin{equation}
    g^{
    (\text{wh,}D)}(\alpha) =  n_D\,\frac{|\Gamma\left(\frac{D-1}{2}+i\alpha\right)|^2}{|\Gamma\left(\frac{1}{2}+i\alpha\right) |^2} \,,\quad\, n_D =  \frac{2^{\frac{5-D}{2}} \pi^{\frac{1+D}{2}}}{\Gamma\left(\frac{D}{2}\right)^2}\,.
\end{equation}
The spin-diagonal first term on the r.h.s.\ of the trace formula gives for this choice of $g$ (or $h$ via \eqref{eq:hdef}) precisely the plateau \eqref{eq:CJplateau}:
\begin{equation}
\label{eq:Naraincorrelations}
\begin{split}
&\frac{\delta_{m_1m_2}}{\pi} \int_\mathbb{R} \frac{d\alpha}{4\pi} \, \alpha  \tanh(\pi \alpha) \, h_{m_1m_2}^{(\text{wh,}D)}(\alpha) = \big\langle {Z}_{\text{P}(D)}^{m_1} {Z}_{\text{P}(D)}^{m_2} \big\rangle_\text{(wh)}
\end{split}
\end{equation}
The spectral overlap coefficients for the $SL(2,\mathbb{Z})$ decomposition of the Narain CFT plateau can simply be read off from the trace formula:
\begin{equation}
    \begin{split}
    \big\langle  z_{\frac{1}{2}+i\alpha_1} \, z_{\frac{1}{2}+i\alpha_2}  \big\rangle_{(\text{wh,}D)} 
   & = \frac{n_D}{\pi} |\Gamma\left(\tfrac{D-1}{2}+i\alpha_1\right)|^2\; 4\pi \delta(\alpha_{12}) \\
    \big\langle z_{n_1} \, z_{n_2} \big\rangle_{(\text{wh,}D)}
  &= \frac{n_D}{\pi} |\Gamma\left(\tfrac{D-1}{2}+iR_{n_1}\right)|^2\, \frac{\delta_{n_1n_2} }{|\!|\nu_{n_1}|\!|^2} 
 \end{split}
 \label{eq:zzNarain}
\end{equation}
The spin-offdiagonal remainder term in the trace formula scales as ${\cal G}_{m_1m_1}^{+(\text{wh,}D)} \sim e^{-2\pi(|m_1|y_1 + |m_2|y_2)}$ for large $y_k$ by the same reasoning described in  Section \ref{sec:specdecomp}. It therefore gives a subleading contribution to the SFF, which is suppressed by an additional factor $\beta^{-D/2}$ relative to \eqref{eq:SFFnarain}.

\subsection{Higher spin theories}
\label{sec:higherspin}

As a final example, consider a 2d CFT with ${\cal W}_N$ symmetry, $Z_{{\cal W}_N}$, dual to higher spin gravity realized as an $SL(N,\mathbb{R})$ Chern-Simons theory. These theories have known unphysical features, e.g., they violate the chaos bound on the Lyapunov exponent \cite{Maldacena:2015waa,Perlmutter:2016pkf}. The SFF for these models was studied in \cite{Das:2021shw} (see also \cite{Kruthoff:2022voq}) and similarly violates random matrix universality. We briefly review this feature in light of the trace formula.

The definition of the ${\cal W}_N$ primary counting partition function involves the removal of $(N-1)$ free bosons:
\begin{equation}
    Z_{\text{P}(N)} =  y^{(N-1)/2} |\eta(x+iy)|^{2(N-1)} Z_{{\cal W}_N}\,.
\end{equation}
This parallels the case of the Narain CFTs with $D \rightarrow N-1$.
Indeed, the SFF for $Z_{\text{P}(N)}$ is of exactly the same form as in the Narain ensemble, \eqref{eq:CJplateau}, however with the replacement $D \rightarrow 2(N-1)$:
\begin{equation}
    \langle {Z}_{\text{P}(N)}^{m_1} {Z}_{\text{P}(N)}^{m_2} \rangle = \left[\langle {Z}_{\text{P}(D)}^{m_1} {Z}_{\text{P}(D)}^{m_2} \rangle \right]_{D \rightarrow 2(N-1)} 
\end{equation}
The different identifications of $D$ in terms of $N$ lead to a different asymptotic behavior in higher spin theories:
\begin{equation}
    \begin{split}
    &(y_1y_2)^{-\frac{N-1}{2}} \langle {Z}_{\text{P}(N)}^{m_1}(y_1) \,{Z}_{\text{P}(N)}^{m_2} ( y_2) \rangle \big{|}_{y_{1,2} \rightarrow \beta \pm iT} \\
    &\quad \sim \text{const.} \times  \beta^{-\frac{D}{2}} \, e^{-4\pi|m_1| \beta} \, T^{N-1}\,\delta_{m_1m_2}\qquad (T \gg \beta) \,.
    \end{split}
    \label{eq:HSramp}
\end{equation}
For $N\geq 3$ this power law growth in $T$ is not consistent with the universal ramp expected for quantum chaotic theories. In this sense, the corresponding spectral overlap coefficients (i.e., \eqref{eq:zzNarain} with $D \rightarrow 2(N-1)$) violate spectral universality. 
The spin-mixing remainder term ${\cal G}_{m_1m_2}^{+(\text{wh,}N)}$ was calculated for $N=3$ in \cite{Das:2021shw} and matches the prediction from the trace formula. As previously, it has additional polynomial suppression in $y_1y_2$ and is thus subleading compared to \eqref{eq:HSramp}.

\section{Discussion}
\label{sec:discussion}

Our analysis concerns general chaotic CFTs, i.e., CFTs whose spectral form factor exhibits universal level repulsion (a `linear ramp') at late times in all spin sectors. We have illustrated that both the completion of this `bare' ramp into a modular invariant SFF as well as their combined modular invariant spectral decomposition are naturally implied and explained by the Kuznetsov trace formula. While the subleading corrections are theory-dependent, we quantified the sense in which the $\mathbb{T}^2\times I$ wormhole amplitude in AdS$_3$ pure gravity is the simplest SFF describing random matrix statistics in CFTs. This emphasizes the universality of the gravity result beyond holography. Any other contributions to a consistent SFF will either give subleading corrections to every term of the trace formula, or will amount to non-trace terms. It is clearly of interest to explore these cases further and characterize quantum chaos beyond the linear ramp.

Our findings streamline some recent discoveries and illuminate the highly constrained interplay between the assumption of quantum chaos and modular invariance. They furthermore manifest that this interplay has deep connections to analytic number theory, thus introducing new concepts and powerful tools into the study of CFTs.

To show how the trace formula captures the spectral decomposition of other SFFs that arise from Poincar\'{e} sums over suitably modular invariant seed functions, we contrasted the linear ramp to examples which either have no ramp (Narain CFTs) or a power law ramp (higher spin theories). In these cases the spectral decomposition is still captured by the trace formula (i.e., it is diagonal both in $SL(2,\mathbb{Z})$ eigenvalues and the functional form of spectral correlations), but the trace part encoded by $h(\alpha)$ is more complicated than for the bare ramp and pure gravity.\footnote{Another case that would be interesting to investigate in this language is the `string' partition function of \cite{DiUbaldo:2023hkc}.}
In these cases the first term on the r.h.s.\ of the trace formula \eqref{eq:kuznetsov} still dominates over the Kloosterman term ${\cal G}^+_{m_1m_2}$, which always correlates different spin sectors. This illustrates an important point: the connection between CFTs and random matrix universality, independently for each spin sector, can only be asserted in the near-extremal limit, where the term diagonal in spin dominates. The trace formula completes this information into a modular invariant object.

\begin{acknowledgments}
The authors thank S.\ Collier and E.\ Perlmutter for helpful conversations and comments. FH is supported by the UKRI Frontier Research Guarantee [EP/X030334/1].  MR and WR are supported by a Discovery Grant from NSERC.
\end{acknowledgments}

\bibliographystyle{apsrev}


\newpage
\appendix

\onecolumngrid

\section*{Supplementary Material}

%

\subsection{Odd parity cusp forms}
\label{app:odd}

In this appendix we generalize the discussion of the ramp for $m\neq 0$ to cusp forms of odd parity. For parity-invariant CFTs, the parity-odd cusp forms form their own superselection sector, which exhibits eigenvalue repulsion and a linear ramp separately from the parity-even sector. To describe both cases, we introduce a parity label `$\pm$' and attach it whenever relevant. We extend the basis of cusp forms to include both of the following:
\begin{equation}
\nu_{n,\pm} = \sum_{m\geq 1}  a_m^{(n,\pm)} \, \left\{ \begin{aligned} \cos(2\pi m x) \\ \sin(2\pi m x) \end{aligned} \right\} \, \sqrt{y} K_{iR_n^\pm} (2\pi m y)\,.
\end{equation}
The eigenvalues of the Laplacian on the upper half plane are $\frac{1}{4}+(R_n^\pm)^2$ for even and odd forms, respectively, where $R_n^+ \gtrsim 13.7$ and $R_n^- \gtrsim 9.5$ are independent sets of sporadic numbers. When including odd cusp forms, we define Fourier coefficients with a sign: $a_{-m}^{(n,\pm)} = \pm a_{m}^{(n,\pm)}$. Similarly, the fixed spin components of the cusp forms, which enter in the trace formula, inherit this sign: $\nu_{n,\pm}^{-m} = \pm  \nu_{n,\pm}^{m}$.

For the even part of the spectrum, the Kuznetsov trace formula was stated in \eqref{eq:kuznetsov}. For the odd part of the spectrum, the analogous trace formula misses the Eisenstein series and reads as follows (for $m_k\neq 0$):\footnote{See, e.g., \cite{motohashi1997} and take the difference of Thm.\ 2.2 and Thm.\ 2.4.}
\begin{equation}
\label{eq:kuznetsovOdd}
    \begin{split}  
        \sum_{n\geq 1} \frac{a_{|m_1|}^{(n,-)}a_{|m_2|}^{(n,-)}}{L_{\nu\times\nu}^{(n,-)}(1)} \, h(R_n^-)  = \frac{\delta_{m_1m_2}}{\pi} \int_{-\infty}^\infty \frac{d\alpha}{4\pi}  \,\alpha  \tanh(\pi \alpha) h(\alpha) + {\cal G}^-_{m_1m_2}\,,
    \end{split}
\end{equation}
where
\begin{equation}
\label{eq:KdefOdd}
\begin{split}
    {\cal G}^-_{m_1m_2} &=2i\, \sum_{c\geq 1} \frac{S(|m_1|,|m_2|;c)}{c} \int_{-\infty}^\infty \frac{d\alpha}{4\pi} \, \frac{\alpha\,h(\alpha)}{\cosh(\pi \alpha)} \, J_{2i\alpha}\left( \frac{4\pi \sqrt{|m_1m_2|}}{c} \right)  \\
    &\quad - \frac{4}{\pi} \,\sum_{c\geq 1} \frac{S(|m_1|,-|m_2|;c)}{c} \int_{-\infty}^\infty \frac{d\alpha}{4\pi} \, \alpha \, h(\alpha)\sinh(\pi \alpha) \, K_{2i\alpha}\left( \frac{4\pi \sqrt{|m_1m_2|}}{c} \right)\,.
\end{split}
\end{equation}

From the discussion in the main text it is immediately clear that the first term on the r.h.s.\ of \eqref{eq:kuznetsovOdd} will also produce a `bare' ramp for the function $h^\text{(wh)}(\alpha)$ chosen as in \eqref{eq:hwh}. The correlations of odd cusp forms which achieve this can again be read off from the trace formula, using the value of the $L$-function given in \eqref{eq:Leval1} (which is identical for odd cusp forms):
\begin{equation}
\label{eq:summaryRampOdd}
 \big\langle z_{n_1,\pm} \, z_{n_2,\pm} \big\rangle_{\text{(wh)}} =  \frac{1}{2\,\cosh(\pi R_{n_1}^\pm)} \, \frac{1}{|\!|\nu_{n_1,\pm}|\!|^2}\times\delta_{n_1n_2}\,.
\end{equation}

We see that, if odd cusp forms are included in the trace, then the universal ramps for spins $m\neq 0$ are produced both from the even and the odd cusp forms. There is a separate trace formula for the ramp in each of the sectors, \eqref{eq:kuznetsov} and \eqref{eq:kuznetsovOdd}. The subleading correction ${\cal G}^-_{m_1m_2}$ is identical to the case of even cusp forms up to a relative sign.

In order to match exactly to the analysis of \cite{Cotler:2020ugk}, one needs to add the even and odd trace formulas (if $\text{sgn}(m_1m_2)=1$) or form their difference (if $\text{sgn}(m_1m_2)=-1$), which simplifies as follows:
\begin{equation}
\begin{split}
&\int_{\mathbb{R}} \frac{d\alpha}{4\pi} \, \frac{4\,a_{m_1}^{(\alpha)} a_{m_2}^{{(\alpha)}}}{|\Lambda(i\alpha)|^2} \, \frac{h(\alpha)}{16\cosh(\pi\alpha)} + \sum_{\epsilon=\pm} \sum_{n\geq 1} \frac{a_{m_1}^{(n,\epsilon)}a_{m_2}^{(n,\epsilon)}}{ |\!|\nu_{n,\epsilon}|\!|^2} \, \frac{h(R_n^\epsilon)}{16\cosh(\pi R_n^\epsilon)}  = \frac{\delta_{m_1m_2}}{\pi} \int_{\mathbb{R}}\frac{d\alpha}{4\pi}  \,\alpha  \tanh(\pi \alpha) h(\alpha) + {\cal G}_{m_1m_2} \,,
\end{split}
\label{eq:generalKuznetsov}
\end{equation}
where 
\begin{equation}
\label{eq:Gsumdiff}
 {\cal G}_{m_1m_2} \equiv \frac{1}{2}\left( {\cal G}^+_{m_1m_2} + \text{sgn}(m_1m_2) \, {\cal G}^-_{m_1m_2} \right) \,.
\end{equation}

\subsection{Generalizations and subleading terms in the wormhole amplitude}
\label{app:subleading}

In this appendix we consider the subleading Kloosterman term ${\cal G}_{m_1m_2}^\text{(wh)}$ and show in more detail that it agrees with the prediction of the Kuznetsov formula. We find it convenient to work with a variant of the latter, which we introduce first.

\subsubsection{Pre-Kuznetsov formula}
\label{app:prekuznetsov}

Consider a function $k: \, \mathbb{H} \times \mathbb{H} \rightarrow \mathbb{R}$, which only depends on the hyperbolic distance between $\tau_1$ and $\tau_2$, i.e., there is an identification
\begin{equation}
   k(\tau_1,\tau_2) \equiv k(u) \equiv k\left( \frac{|\tau_1-\tau_2|^2}{4y_1y_2} \right) \,.
\end{equation} 
This property is guaranteed, for example, if $k$ is invariant under joint $SL(2,\mathbb{R})$ transformations, $k(\gamma \tau_1, \, \gamma \tau_2) = k(\tau_1,\tau_2)$.\footnote{ A sufficient condition would also be that $k$ takes the form of a Poincar\'{e} sum over $SL(2,\mathbb{Z})$ images of a seed function with certain diagonal pairing of eigenvalues under Hecke operators \cite{DiUbaldo:2023qli}. We thank E.\ Perlmutter for pointing this out.}

We define the {\it Selberg transform} $\hat{k}(\alpha)$ through
\begin{equation}
\begin{split}
   \hat{k}(\alpha) &\equiv   \int_{\mathbb{H}} \frac{dx dy}{y^2} \,y^{\frac{1}{2} + i\alpha} \,  {k} (i,x+iy)  \,,\\
   k(u) &= \int_\mathbb{R} \frac{d\alpha}{4\pi} \, \alpha \tanh(\pi \alpha) \, P_{-\frac{1}{2}+i\alpha}(2u+1)\, \hat{k}(\alpha) \,,
\end{split}
\end{equation}
where the Legendre function is
\begin{equation}
    P_{-\frac{1}{2}+i\alpha}(2u+1) 
    \equiv \frac{1}{\pi}\int_0^\pi d\theta \, \big( 2u+1+2\sqrt{u(u+1)} \, \cos\theta \big)^{-\frac{1}{2}+i\alpha} \,.
\end{equation}

Under similar assumptions as for the Kuznetsov formula, the {\it pre-Kuznetsov trace formula} is stated in terms of $k(\tau_1,\tau_2)$ and its Selberg transform (see, e.g., \cite{Duke2002,Li}):
\begin{equation}
\label{eq:prekuz}
\begin{split}
   &\int_\mathbb{R}\frac{d\alpha}{4\pi}\, \hat{k}(\alpha) \, E^{m_1}_{\frac{1}{2}+i\alpha}(y_1) \, E^{m_2}_{\frac{1}{2}+i\alpha}(y_2)
   + \sum_{\epsilon=\pm} \sum_{n \geq 1}\hat{k}(R_n^\epsilon)\,  \frac{\nu_{n,\epsilon}^{m_1}(y_1)}{|\!|\nu_{n,\epsilon}|\!|} \frac{\nu_{n,\epsilon}^{m_2}(y_2)}{|\!|\nu_{n,\epsilon}|\!|}
   \\
   &\quad = 4\,\delta_{m_1m_2} \, \int_\mathbb{R} dx \, e^{2\pi i m_1 x_1} k(x_1+iy_1 , i y_2)\\
   &\qquad  + 4 \sum_{c\geq 1} S(m_1,m_2;c) \int_\mathbb{R} dx_1 \int_\mathbb{R} dx_2 \, e^{2\pi i(m_1 x_1 - m_2 x_2)} \, k \left( x_1 + i y_1 , \, \frac{-1}{c^2(x_2+iy_2)}\right)\,,
\end{split}
\end{equation}
where we assume sgn$(m_1m_2)=1$ for definiteness. Note that \eqref{eq:prekuz} corresponds to the sum of \eqref{eq:kuznetsov} and \eqref{eq:kuznetsovOdd}, which is valid for spins of equal signs and contains cusp forms of both parities. 
Being manifestly a spectral decomposition, this formula is obviously useful for our purposes.
It is straightforward to recover the Kuznetsov formula \eqref{eq:kuznetsov} from the above \cite{Duke2002,Li}. In particular, we identify $\hat{k}(\alpha) = g(\alpha)/(4\cosh(\pi\alpha))$.

Comparing with \eqref{eq:generalKuznetsov}, the case of the pure gravity wormhole (or `bare' linear ramp) evidently corresponds to the choice 
\begin{equation}
 \hat{k}^\text{(wh)}(\alpha) = \frac{1}{4 \cosh(\pi \alpha)} \,.
\end{equation}
The associated function $k^\text{(wh)}(u) \equiv k^\text{(wh)}(\tau_1,\tau_2)$ entering on the geometric side of the pre-Kuznetsov formula is
\begin{equation}
    \begin{split}
    k^\text{(wh)}(u) &= \frac{1}{4} \int_\mathbb{R} \frac{d\alpha}{4\pi} \, \frac{\alpha \tanh(\pi \alpha)}{\cosh(\pi \alpha)} \, P_{-\frac{1}{2}+i\alpha}(2u+1) \\
    &= \frac{1}{16\pi^3} \int_0^\pi d\theta \; \frac{2(1+X)+(1-X)\log(X)}{(1+X)^2} \qquad \left[ X\equiv 2u + 1 + 2\sqrt{u(u+1)} \, \cos \theta \right] \\
    &= \frac{1}{16\pi^2} \, \frac{1}{1+u} \,.
    \end{split}
\end{equation}
The spin-diagonal term on the r.h.s.\ of the pre-Kuznetsov formula is the `bare' ramp, i.e.,
\begin{equation}
    \begin{split}
    4\,\delta_{m_1m_2} \int dx \, e^{2\pi i m_1 x_1} k^\text{(wh)}(x_1+iy_1 , i y_2)
    &=  \frac{\delta_{m_1m_2}}{\pi} \frac{y_1 y_2}{y_1+y_2} \, e^{-2\pi |m_1| (y_1+y_2)} \,. 
    \end{split}
\end{equation}

\subsubsection{Matching Kloosterman terms}

We will now discuss in more detail the subleading term ${\cal G}_{m_1m_2}^\text{(wh)}$ in the Kuznetsov trace formula, see \eqref{eq:Gsumdiff}. 
We wish to show that it equals the known gravity result, i.e., $\pi^{-2}$ times eq.\ (4.20) of \cite{Cotler:2020ugk}:\footnote{Compared to \cite{Cotler:2020ugk}, we have a different sign in one term in the exponent. We believe this to be a typo in that reference.}
\begin{equation}
\begin{split}
 {\cal G}^\text{(wh,\cite{Cotler:2020ugk})}_{m_1m_2} 
  &= \frac{1}{\pi}
  \sum_{c\geq 1} \frac{S(m_1,m_2;c)}{ c^2 B_c}\,e^{- 2\pi \, \text{sgn}(m_1) \left(m_1 y_1-m_2 y_2 \right)} \int_{-\infty}^\infty \frac{dx}{x^2+1} \, e^{- 2\pi  m_2 y_2   \left(\text{sgn}(m_1) + i B_c x\right)- \frac{2\pi  m_1}{c^2  y_2 \left(\text{sgn}(m_1)-i B_c x \right)} }  \,.
\end{split}
\label{eq:CJfinal}
\end{equation}
Note that, if sgn$(m_1m_2)=-1$, then the $x$-integral can be performed by deforming the contour of integration and evaluating the residue at $x= \pm i$ (for sgn$(m_1)=\mp1$, respectively); the result matches \eqref{eq:GwhOpposite} or the combination appearing in \eqref{eq:generalKuznetsov} for $h^\text{(wh)}(\alpha)$.

To discuss spins with equal signs, it is convenient to use the pre-Kuznetsov formula \eqref{eq:prekuz}.
The second term on its r.h.s.\ is the Kloosterman term. For the wormhole amplitude, the relevant integral is:
\begin{equation}
    \begin{split}
     {\cal G}_{m_1m_2}^\text{(wh)} &= 4 \sum_{c\geq 1} S(m_1,m_2;c) \int_\mathbb{R} dx_1 \int_\mathbb{R} dx_2 \, e^{2\pi i(m_1 x_1 - m_2 x_2)} \, k^\text{(wh)} \left( x_1 + i y_1 , \, \frac{-1}{c^2(x_2+iy_2)}\right) \\
     &= \frac{1}{\pi} \sum_{c\geq 1} S(m_1,m_2;c) \int_\mathbb{R} dx_2  \, e^{-2\pi i m_2 x_2} \, \frac{y_1 y_2}{y_2 + c^2 y_1(x_2^2 + y_2^2)} \, e^{-2\pi |m_1| \left( y_1 - \frac{1}{ic^2(\text{sgn}(m_1)\,x_2+i y_2)} \right)} \\
     &= \frac{1}{\pi}\sum_{c\geq 1} \frac{S(m_1,m_2;c)}{ c^2 B_c} \, e^{- 2\pi |m_1| y_1}\int_\mathbb{R} \frac{dx}{1+x^2} \, e^{-2\pi i m_2 y_2 B_c x - \frac{2\pi |m_1|}{c^2 y_2 (1 - i\, \text{sgn}(m_1)\,B_c x)} }
    \end{split}
\end{equation}
This matches precisely the subleading terms found in the gravity analysis, \eqref{eq:CJfinal}. This match is by design, since \cite{Cotler:2020ugk} derives the wormhole amplitude as a Poincar\'{e} sum over images of a seed function $\mathbb{H} \times \mathbb{H} \rightarrow \mathbb{R}$ that is invariant under joint modular transformations and the above formalism hence applies to it. One subtlety to note is that \cite{Cotler:2020ugk} actually consider seeds satisfying a twisted modular invariance condition, $k(\gamma\tau_1,\, \gamma^{-1} \tau_2) = k(\tau_1,\tau_2)$ for $\gamma \in SL(2,\mathbb{Z})$. For the purpose of showing a match between terms, this is irrelevant; but, of course, the formalism described above could easily be adapted to this setup. 

The same formalism can also be used to confirm that the Kloosterman terms predicted by the trace formula for CFTs with ${\cal W}_N$ symmetry match those found in \cite{Das:2021shw}.

\end{document}